\documentclass[10pt, a4paper, conference]{IEEEtran}
\IEEEoverridecommandlockouts
\usepackage{cite}
\usepackage{amsmath,amssymb,amsfonts}
\usepackage{algorithmic}
\usepackage{graphicx}
\usepackage{textcomp}
\usepackage{xcolor}
\usepackage{multirow}
\usepackage{bm}
\def\BibTeX{{\rm B\kern-.05em{\sc i\kern-.025em b}\kern-.08em
    T\kern-.1667em\lower.7ex\hbox{E}\kern-.125emX}}
\begin{document}

\title{Anomaly Detection in Time Series of EDFA Pump Currents to
Monitor Degeneration Processes using Fuzzy Clustering\\
\thanks{This work is supported by the SNS Joint Undertaken under grant agreement No. 101096120 (SEASON). Responsibility for the content of this publication is with the authors.

\copyright2024 IEEE. Personal use of this material is permitted. Permission from IEEE must be obtained for all other uses, in any current or future media, including reprinting/republishing this material for advertising or promotional purposes, creating new collective works, for resale or redistribution to servers or lists, or reuse of any copyrighted component of this work in other works.}
}

\author{\IEEEauthorblockN{Dominic Schneider}
\IEEEauthorblockA{\textit{Advanced Technology} \\
\textit{Adtran}\\
98617 Meiningen, Germany \\
dominic.schneider@adtran.com}
\and
\IEEEauthorblockN{Lutz Rapp}
\IEEEauthorblockA{\textit{Advanced Technology} \\
\textit{Adtran}\\
98617 Meiningen, Germany\\
lutz.rapp@adtran.com}
\and
\IEEEauthorblockN{Christoph Ament}
\IEEEauthorblockA{\textit{Faculty of Applied Computer Science} \\
\textit{University of Augsburg}\\
86159 Augsburg, Germany \\
christoph.ament@uni-a.de}
}

\maketitle

\begin{abstract}
This article proposes a novel fuzzy clustering based anomaly detection method for pump current time series of EDFA systems. The proposed change detection framework (CDF) strategically combines the advantages of entropy analysis (EA) and principle component analysis (PCA) with fuzzy clustering procedures. In the framework, EA is applied for dynamic selection of features for reduction of the feature space and increase of computational performance. Furthermore, PCA is utilized to extract features from the raw feature space to enable generalization capability of the subsequent fuzzy clustering procedures. Three different fuzzy clustering methods, more precisely the fuzzy clustering algorithm, a probabilistic clustering algorithm and a possibilistic clustering algorithm are evaluated for performance and generalization. Hence, the proposed framework has the innovative feature to detect changes in pump current time series at an early stage for arbitrary points of operation, compared to state-of-the-art predefined alarms in commercially used EDFAs. Moreover, the approach is implemented and tested using experimental data. In addition, the proposed framework enables further approaches of applying decentralized predictive maintenance for optical fiber networks.
\end{abstract}


\section{Introduction}
Erbium--doped fiber amplifiers (EDFAs) are key elements in current long--haul optical fiber transmission networks. Since even short--term failures of an optical link cause cost--intensive loss of transmission capacity, high reliability of the installed EDFAs is of major importance. Core components of such amplifiers enabling amplification of optical data signals are erbium--doped fibers (EDFs) and pump lasers used to excite the erbium ions. A soft--failure of the pump laser caused by aging and degeneration processes will lead to performance degradation of the whole system.

Nowadays, commercially used EDFAs are typically operated with an integrated automatic gain control (AGC) stabilizing the amplifier gain by controlling the pump current and thus the pump power. Aging of the pump lasers leads to either sudden death or slow reduction of pump power at constant pump current due to wear--out that can be detected by monitoring the pump current and the emitted pump power over time. When the pump current required to obtain the maximum output power exceeds a threshold, an alarm raised.

Standard techniques determine whether the pump current required to generate the maximum pump power exceeds a pre--defined threshold. However, aging detection does not work reliably when operating an optical amplifier below its maximum output power. In this paper, we propose a change detection framework to detect anomalies in the pump current time series allowing to detect aging of the pump at arbitrary points of operation and thus to reduce the size of the dataset provided to the evaluation algorithm. The proposed method is based on fuzzy clustering and is analyzed by means of experimental data.

\section{Fuzzy Clustering Procedures}
Many different methods are known to detect anomalies in time series. The most common methods are based on stochastic time series analysis, mathematical models, classification, clustering, etc. \cite{GOMBAY2008451} \cite{Aminikhanghahi2017}. The task becomes even more complicated as the changes develop in bounds in a non--stationary environment. With smoothly and slowly developing changes in the time series, the definition of crisp sets is not feasible, whereas fuzzy clustering methods showed reliable learning and generalization behavior in various applications \cite{bezdek2013pattern}.

The aim of clustering is to identify data samples belonging to the same homogenous clouds of observations comprising similar data points. Such a clustering of real--world data is impaired by uncertainties caused by noise, missing values, etc which are forming "smeared" boundaries, so--called fuzzy sets. For this reason, the classical fuzzy c--means (FCM) algorithm was developed, as mathematically represented in Eq. (\ref{eq:C-Means Clustering Algorithm}) \cite{bezdek2013pattern}
\begin{equation}\label{eq:C-Means Clustering Algorithm}
   \left\{ \begin{array}{l}
   \bm{w_{j}(k)}=\frac{\|\mathbf{x(k)}-\mathbf{c_{j}}\|^{-2}}{\sum_{l=1}^{m}\|\mathbf{x(k)}-\mathbf{c_{l}}\|^{-2}}\\[10pt]
   \mathbf{c_{j}}=\frac{\sum_{k=1}^{N}\bm{w_{j}^{2}(k)}\mathbf{x(k)}}{\sum_{k=1}^{N}\bm{w_{j}^{2}(k)}}
   \end{array}\right.,
\end{equation}
where $\bm{w_{j}(k)}$ denotes the weight {\it j} of data sample {\it k}. It is calculated as normalized Euclidian distance of the {\it n}-dimensional vector of features $\mathbf{x(k)}$ from the data sample {\it k} to the {\it n}-dimensional center $\mathbf{c_{j}}$ of the cluster {\it j}. Therefore, $\bm{w_{j}(k)}$ is a measurement of the affinity of a data sample to a cluster. As the algorithm aims to find the best fit of centers for all clusters, an update of the initial clusters need to be performed by weighting the data sample $\mathbf{x(k)}$ with the quadrating of $\bm{w_{j}(k)}$ and normalizing over all data samples.

In the traditional approach, a data sample is assigned to a cluster with a crisp set. This means that the data sample is either in the cluster or not. In the case of a fuzzy cluster analysis, the data sample is assigned to a cluster with a membership value between in $[0, 1]$. These methods developed into two main directions: probabilistic and possibilistic clustering approaches. In \cite{Bodyanskiy2019}, the authors propose to use both approaches for the task of anomaly detection in time series. In Eq. (\ref{eq:Distance}), a robust distance function $D^{R}(\mathbf{x(k)},\mathbf{c_{j}})$ is defined, based on the standard activation function of an artificial neuron
\begin{equation}\label{eq:Distance}
   D^{R}\left(\mathbf{x(k)},\mathbf{c_{j}}\right)=\sum_{i=1}^{n}\bm{\beta_{i}}\ln\left[\cosh\left(\frac{\bm{x_{i}(k)}-\bm{c_{ji}}}{\bm{\beta_{i}}}\right)\right],
\end{equation}
and defines a scalar metric for the distance between a data sample $\mathbf{x(k)}$ and a cluster center $\mathbf{c_{j}}$. It utilizes the hyperbolic cosine in combination with a natural logarithm to reduce the amplification of outliers, which are far away from the center of a cluster. Summarizing over all features {\it n} results in one metric per data sample to each cluster center, similar to the weights in Eq. (\ref{eq:C-Means Clustering Algorithm}). The parameter $\bm{\beta_{i}}$ controls the speed of change of the function and serves as a hyperparameter for further regularization.

Introducing the metric of Eq. (\ref{eq:Distance}) and utilizing it in the probabilistic clustering procedure (ProbCP), the following algorithm is defined by Eq. (\ref{eq:Robust Probabilistic Clustering Algorithm})
\begin{equation}\label{eq:Robust Probabilistic Clustering Algorithm}
  \left\{ \begin{array}{l}
   \bm{w_{j}^{pr}(k)}=\frac{\left(D^{R}\left(\mathbf{x(k)}-\mathbf{c_{j}}\right)\right)^{\frac{1}{1-\beta}}}{\sum_{l=1}^{m}\left(D^{R}\left(\mathbf{x(k)}-\mathbf{c_{j}}\right)\right)^{\frac{1}{1-\beta}}}\\[10pt]
   \begin{aligned}
      \bm{c_{ji}^{pr}(k+1)}=\bm{c_{ji}^{pr}(k)}+\eta(k)\bm{w_{j}^{\beta}(k)} \\
      \tanh\left(\frac{\bm{x_{i}(k)}-\bm{c_{ji}(k)}}{\bm{\beta_{i}}}\right)
   \end{aligned}
   \end{array}\right.,
\end{equation}
where $\bm{w_{j}^{pr}(k)}$ serves as metric for the membership of a data sample $\mathbf{x(k)}$ to an existing cluster center $\mathbf{c_{j}^{pr}}$, which is similar to the classic fuzzy clustering represented in Eq. (\ref{eq:C-Means Clustering Algorithm}), whereas the distance is calculated using Eq. (\ref{eq:Distance}) rather than the Euclidian norm. The exponent $\frac{1}{1+\beta}$ influences the shape of the distance function and represents an additional hyperparameter for regularization. In contrast to Eq. (\ref{eq:C-Means Clustering Algorithm}), the update of $\bm{c_{ij}^{pr}(k)}$ in Eq. (\ref{eq:Robust Probabilistic Clustering Algorithm}) is carried out iteratively. The prior value for the cluster center is accumulated with a gradient term, whereas $\eta(k)$ denotes the learning rate of the algorithm and the weighted hyperbolic tangent resembles a gradient value. The learning rate $\eta(k)$ can be adapted iteratively to reduce eventually occurring oscillating behavior of the algorithm.
\begin{figure}[!t]
   \centering
        \includegraphics[width=\linewidth]{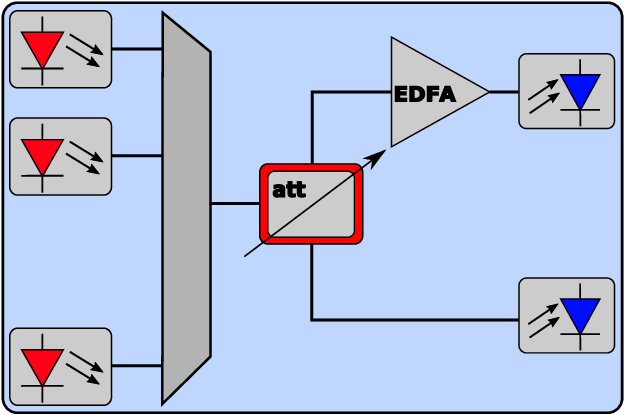}
    \caption{Data acquisition setup}
    \label{fig:dataAS}
\end{figure}

In addition to the use of weights, the possibilistic clustering procedure (PossCP) makes use of already observed data samples for building the clusters. This results in a new metric $\bm{\mu_{j}}$ and a defined algorithm by Eq. (\ref{eq:Robust Possibilistic Clustering Algorithm})
\begin{equation}\label{eq:Robust Possibilistic Clustering Algorithm}
   \left\{ \begin{array}{l}
   \bm{w_{j}^{pos}(k)}=\left(1+\left(\frac{D^{R}\left(\mathbf{x(k)},\mathbf{c_{j}}\right)}{\bm{\mu_{j}}}\right)^{\frac{1}{\beta-1}}\right)^{-1}\\[10pt]
   \begin{aligned}
      \bm{c_{ji}^{pos}(k+1)}=\bm{c_{ji}^{pos}(k)}+\eta(k)\bm{w_{j}^{\beta}(k)} \\
      \tanh\left(\frac{\bm{x_{i}(k)}-\bm{c_{ji}(k)}}{\bm{\beta_{i}}}\right)
   \end{aligned}\\[10pt]
   \bm{\mu_{j}(k+1)}=\frac{\sum_{p=1}^{k}\bm{w_{j}^{\beta}(p)}D^{R}\left(\mathbf{x(p)},\mathbf{c_{j}(k+1)}\right)}{\sum_{p=1}^{k}\bm{w_{j}^{\beta}(p)}}
   \end{array}\right.,
\end{equation}
where the weights $\bm{w_{j}^{pos}(k)}$ are calculated utilizing the predefined distance function of Eq. (\ref{eq:Distance}) and an exponential implementation of a hyperparameter $\frac{1}{\beta-1}$ for regularization, whereas the distance function is normalized by $\bm{\mu_{j}}$. This scalar metric observes the already presented data samples and adapts dynamically the distance at which the membership level takes the value $0.5$, which results in $\bm{w_{j}^{pos}(k)}=0.5$. The iterative update of the cluster centers is similar to the probabilistic clustering procedure, represented in Eq. (\ref{eq:Robust Probabilistic Clustering Algorithm}).

\section{Experimental Setup}

The used data acquisition setup is shown in Fig.~\ref{fig:dataAS}, where a wavelength--division multiplexing (WDM) signal is amplified by a commercial EDFA. The WDM signal contains ten channels, equally distributed in the conventional wavelength band (C--band). The input power level is set to a range from -35\,dBm to 1\,dBm. The gain is set to values ranging from 19\,dB to 35\,dB. Furthermore, the maximum output power is limited to 20\,dBm. With this setup, a total dataset comprising 11,886 samples with 41 features is generated.

The features mainly contain monitoring values from the EDFA. The features are divided into three groups: optical, electrical and temperature readings. First, in the optical part we have the optical input power, the optical output power, the optical input power of the second stage and the optical output power of the first stage. Second, in the electrical part we have monitoring values about the currents and pump powers for the EDFA stages and more operating voltages. Finally, in the temperature part we have the temperature of the EDFA housing and the temperatures of the pump laser chips.

With the aim of detecting anomalies in the time series of pump current data, variations are introduced in these data. Pump degeneration is a well--researched process~\cite{6836254}\cite{Bliznyuk2021}. Due to aging, the pump current required to reach a desired optical pump power starts to increase over time~\cite{9670373}. The ratio of actual pump current to nominal current can be used to model these effects and to create a specific data stream for evaluating the algorithms.

\section{Change Detection Framework}

\begin{figure}[!t]
   \centering
        \includegraphics[width=\linewidth]{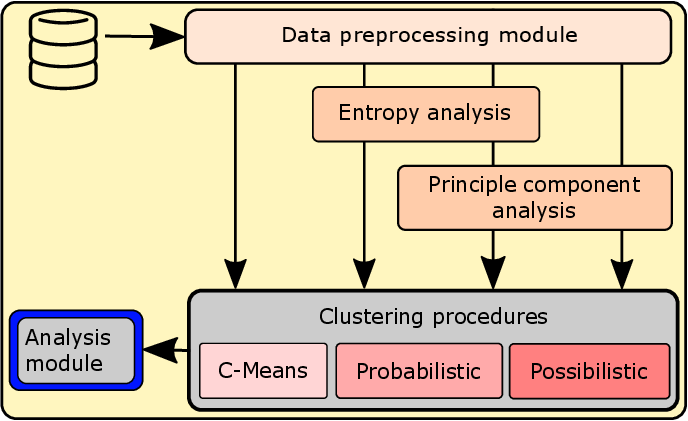}
    \caption{Change detection framework}
    \label{fig:cdFramework}
\end{figure}

Artificial intelligence (AI) models can suffer from the curse of dimensionality. This means that the amount of features that can be learned by a machine learning algorithm is dependent on the algorithms capacity. Rather than overcoming this problem by using deep learning architectures, we propose a feature selection and reduction process. Combined with the clustering procedures, this approach results in a change detection framework (CDF), which is shown in Fig.~\ref{fig:cdFramework}.

The {\it Data Preprocessing Module} removes irrelevant, not explainable, null or repeated features from the dataset. Machine learning algorithms are sensitive to the scale of the data \cite{9030052}. Therefore this module implements a standard scaler to normalize the data.

{\it Feature Selection} uses a wide variety of methods which measure the impact of different features on the target value. Due to a missing target value in the task of clustering, these algorithms are not suitable \cite{khalid2014survey} for the present problem. Therefore, we propose a method derived from information theory wherein information of a distribution is measured by it's entropy, as represented in Eq. (\ref{eq:Entropy}).
\begin{equation}\label{eq:Entropy}
   H(X)=-\sum_{k=1}^{N}\bm{p_{k}}\ln \bm{p_{k}}\;,
\end{equation}
where the entropy accumulates the weighted probabilities of events of a feature. Through the use of the natural logarithm, the units are called {\it nats}. 
Calculating the entropy for each feature and selecting all usable features with $H(X)_{i} > H_{min}$ and $H_{min} = 0$ results in a suitable feature selection method, whereas no transformations are performed.
\begin{figure}[!t]
   \centering
      \includegraphics[width=0.9\linewidth]{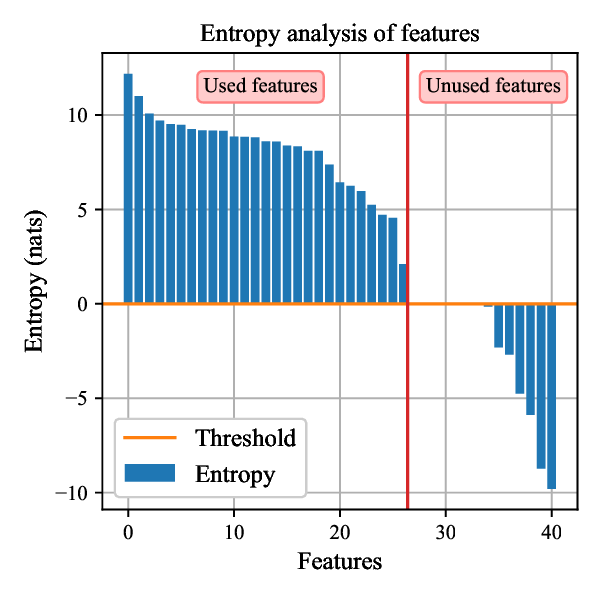}
   \caption{Entropy analysis on feature space}
   \label{fig:entropy}
\end{figure}
\begin{figure}[!b]
   \centering
      \includegraphics[width=0.9\linewidth]{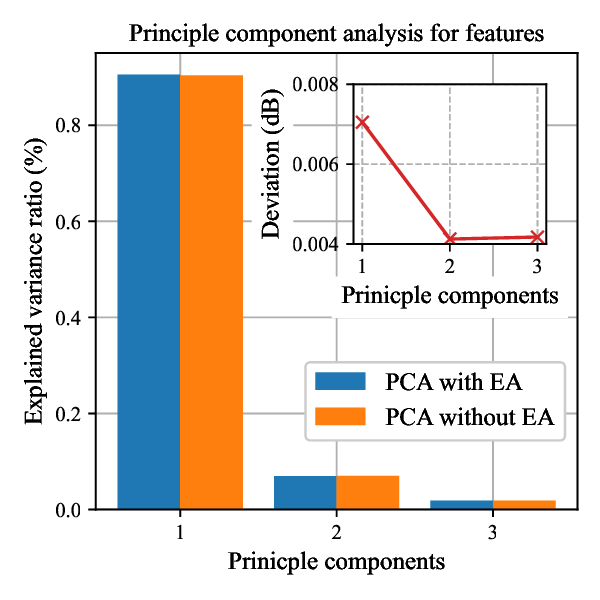}
   \caption{Principle component analysis on feature space}
   \label{fig:pca}
\end{figure}
\begin{figure*}[!t]
   \centering
        \includegraphics[width=\linewidth]{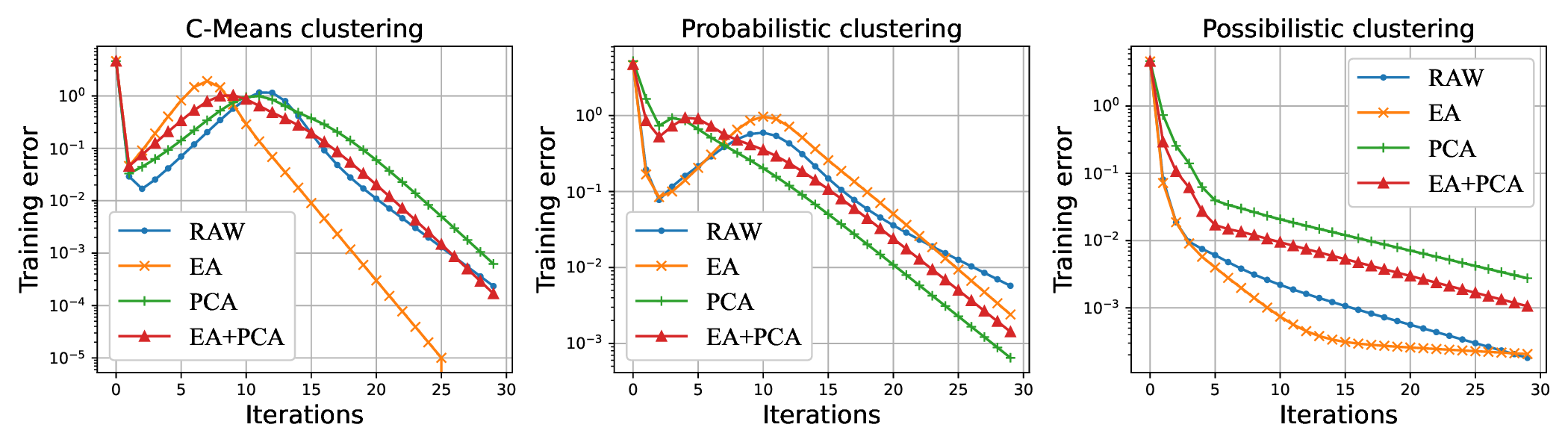}
    \caption{Training Error of Clustering Procedures}
    \label{fig:error}
\end{figure*}
The entropy analysis (EA) is illustrated in Fig.~\ref{fig:entropy}. The initial feature space of 41 features is analyzed for the entropy per feature and sorted. Setting the selecting threshold to $H_{min} = 0$, leads to a reduction of used features, as only features with a higher entropy will be considered in further processing. It is noticeable that 27 features are remaining after applying the EA.

Finally, the {\it Feature Extraction} uses a latent model, the Principle Component Analysis (PCA), to reduce the dimensionality of the data. The PCA applies a linear transformation to the data based on the eigenvectors of the covariance matrix of the data. The eigenvectors with the highest eigenvalues are then used to transform the data \cite{murphy2012machine}. The PCA is applied to two different datasets, to the modified dataset by EA and the raw dataset, which is shown in Fig.~\ref{fig:pca}. The result gives three principle components, due to the setting that the accumulated explained variance ratio need to exceed $95\%$. The explained variance ratio is a metric of how much variance each principle component contains regarding the original database. The first principle component exceeds the other two principle components far for both databases. The deviation of each principle component, compared by the two datasets, is quite low, but will have a greater effect in generalization, as later shown.

\section{Results and Analysis}
\subsection{Training behavior analysis}
The learning behavior of the algorithms under different settings for the CDF is illustrated in Fig.~\ref{fig:error}. First of all, it is noticeable that the convergence criterium is reached within $30$ iterations, respectively. Nonetheless, the classical FCM tends to soar in the first iterations for all given CDF configurations, which implies trouble in finding the best gradient for the provided dataset. In contrast, the ProbCP reduces the soaring and provides a more robust behavior for the CDF configuration with enabled EA and EA + PCA. Finally, the PossCP eliminates the soaring completely and resembles the most robust learning behavior for all three clustering algorithms and is independent over all CDF configurations.

\subsection{Ablation study of the CDF}
In the experiment, the learning rate is set to a fixed value of $\eta=10^{-3}$. Using $\lVert \mathbf{W^{t+1}}-\mathbf{W^{t}}\rVert\le\epsilon$ with $\epsilon=10^{-4}$, a suitable stop--criterium is created. The hyperparameters of the algorithms are set to $\beta=2$ and $\bm{\beta_{i}}=1$. For the purpose of stochastic variance and the random initialization of the weight matrices, the number of runs equals $n=25$ for averaging the results.

First, the performance of the algorithms are evaluated on the raw data. The results are shown in Tab.~\ref{tab:clustResults}. The mean square error (MSE) serves as metric for the performance of the algorithm and is measured by the training ($MSE_{tr}$) and test ($MSE_{te}$) dataset. The results show that the FCM performs worse (29.3\,\%) in terms of the training dataset as compared with ProbCP (0.0\,\%) and PossCP (0.0\,\%). The evaluation on the test dataset shows that no algorithm is able to generalize, which leads to unsatisfied performance.
\begin{table}[b]
   \centering
   \caption{Results of the clustering algorithms}
   \label{tab:clustResults}
      \begin{tabular}{|l|c|c|c|c|}
         \hline {\textbf{CDF-Config}} & {\textbf{Error-type}} & {\textbf{FCM}} & {\textbf{ProbCP}} & {\textbf{PossCP}} \\
         \hline \multirow{2}{*}{RAW} & $MSE_{tr}$ & 29.3 \% & 0.0 \% & 0.0 \% \\
         \cline{2-5} & $MSE_{te}$ & 99.7 \% & 99.7 \% & 99.7 \% \\
         \hline \multirow{2}{*}{EA} & $MSE_{tr}$ & 17.5 \% & 0.0 \% & 0.0 \% \\
         \cline{2-5} & $MSE_{te}$ & 86.3 \% & 78.7 \% & 74.3 \% \\
         \hline \multirow{2}{*}{PCA} & $MSE_{tr}$ & 15.4 \% & 0.0 \% & 0.0 \% \\
         \cline{2-5} & $MSE_{te}$ & 30.3 \% & 22.7 \% & 18.9 \% \\
         \hline \multirow{2}{*}{PCA+EA} & $MSE_{tr}$ & 12.4 \% & 0.0 \% & 0.0 \% \\
         \cline{2-5} & $MSE_{te}$ & 29.3 \% & 21.3 \% & 17.7 \% \\
         \hline
      \end{tabular}
\end{table}

Second, the performance of the algorithms is evaluated by using the data after the feature selection with EA. The results are shown in Tab.~\ref{tab:clustResults}. The EA slightly improves the performance of the algorithms on the training and test dataset. Nonetheless, the performance of the algorithms on the test dataset with 86.3\,\%, 78.7\,\% and 74.3\,\% is insufficient.

Third, we introduce the PCA to reduce the dimensionality of the dataset and to perform an orthogonal transformations on the features. As shown in Tab.~\ref{tab:clustResults}, the PCA has a positive impact on the performance of the algorithms on both, training and test dataset. The PCA enables the algorithms to generalize from the training dataset and improves the performance on the test dataset up to 30.3\,\%, 22.7\,\%, and 18.9\,\% for the FCM, ProbCP and PossCP, respectively.

Finally, the performance is determined for a combination of EA and PCA which is shown in Tab.~\ref{tab:clustResults}. The EA can improve the performance of the algorithms with PCA up to 3.0\,\% in terms of the training dataset and up to 1.4\,\% in terms of the test dataset. This is due to a change in the principle components, which are generated by the transformation and thus produce better learnability. Therefore the combination of the fuzzy clustering algorithms with the EA and PCA is the best performing setup for the CDF.
\begin{table}[t]
   \centering
   \caption{Results of the clustering algorithms}
   \label{tab:compResults}
      \begin{tabular}{|l|c|c|c|c|}
         \hline {\textbf{Method}} & {$\mathbf{MSE_{tr}}$} & {$\mathbf{MSE_{te}}$} \\
         \hline K-Means clustering & 39.87 \% & 45.33 \% \\
         \hline Hierarchical clustering & 39.32 \% & 47.53 \% \\
         \hline BIRCH & 31.67 \% & 34.17 \% \\
         \hline CDF & 0.0 \% & 17.7 \% \\
         \hline
      \end{tabular}
\end{table}

\subsection{Comparison with other methods}
To evaluate the performance of the CDF, a comparison to current state-of-the-art methods is done. In \cite{nassif2021machine} an overview of available clustering methods in the field of machine learning is given. Therefore, a direct comparison of the CDF with algorithms of the same class is done: 1) K-Means clustering, 2) hierarchical (agglomerative) clustering and 3) Balanced Iterative Reducing and Clustering using Hierarchies (BIRCH) clustering. K-Means clustering was originally invented for signal processing and utilizes vector quantization. Hereby, $n$ observations are clustered into $k$ clusters whereas the nearest mean serves as cluster prototypes. It has been successfully applied to the task of unsupervised anomaly detection in time series as shown in \cite{westa2023unsupervised}. Hierarchical clustering uses a linkage distance, the Euclidian norm or any other distance metric, and recursively merges pairs of clusters to build a classification tree \cite{tokuda2022revisiting}. BIRCH is an online-learning algorithm with a high memory-efficiency. Similar to hierarchical clustering, a tree-like data structure is constructed. Lately, the successful application to detect anomalies in time series with lower dimensions has been proven \cite{guo2023anomaly}. In order to reduce the impact of randomness, the experiment was repeated 20 times and the metrics are averaged. As shown in Tab.~\ref{tab:compResults}, the chosen performance metric for evaluation is MSE for training as well as test scenario. K-Means clustering has a poor learning behavior (39.87\,\%) and is not able to generalize. With a test performance of 45.33\,\%, an oscillating behavior is likely as we have a two-class problem. Hierarchical clustering shows similar results as the K-Means algorithm with a performance of 39.32\,\% in training and 47.53\,\% in test, respectively. BIRCH provides a better performance in training (31.67\,\%) and test (34.17\,\%) than the beforementioned algorithms. Of interest is the significant performance deviation between hierarchical clustering and BIRCH. Both algorithms construct a hierarchical data structure, but in contrast hierarchical clustering has a bottom up approach, whereas each observation starts in it's own cluster. More importantly, the proposed CDF outperforms all compared ones with the smallest values for MSE in training and test.

\subsection{Experimental anomaly detection study}
The purpose of the CDF is to detect anomalies in time series of the pump current, but the performance metrics of the training and test dataset are quantified measurements for shuffled datapoints. In order to evaluate the performance of the algorithms on the time series, we introduce the change--point detection (CPD) as a metric. The CPD is defined as the minimal change in the pump current which is detected by the algorithms with EA + PCA, shown in Tab.~\ref{tab:minCPD}. The value $I_{0}$ denotes the nominal pump current and $I$ the actual pump current. Therefore $I/I_{0}$ can be interpreted as drift. The results show that the C--Means algorithm is able to detect a drift of 8.1\,\% in the pump current. The Probabilistic and Possibilistic algorithms are able to detect a drift of 5.9\,\% and 4.9\,\% in the pump current, respectively. Therefore the PossCP enables the CPD significantly earlier than the predefined thresholds, which are typically set to 10\,\%, especially for arbitrary operating conditions.
\begin{table}[t]
   \centering
\caption{Minimal Change-Point Detection}
\label{tab:minCPD}
      \begin{tabular}{|l|c|c|c|}
         \hline  &  \textbf{FCM} & \textbf{ProbCP} & \textbf{PossCP}\\
         \hline $I/I_{0}$ & 8.1 \% & 5.9 \% & 4.9 \%\\
         \hline
      \end{tabular}
\end{table}
\begin{figure}[!b]
   \centering
        \includegraphics[width=0.9\linewidth]{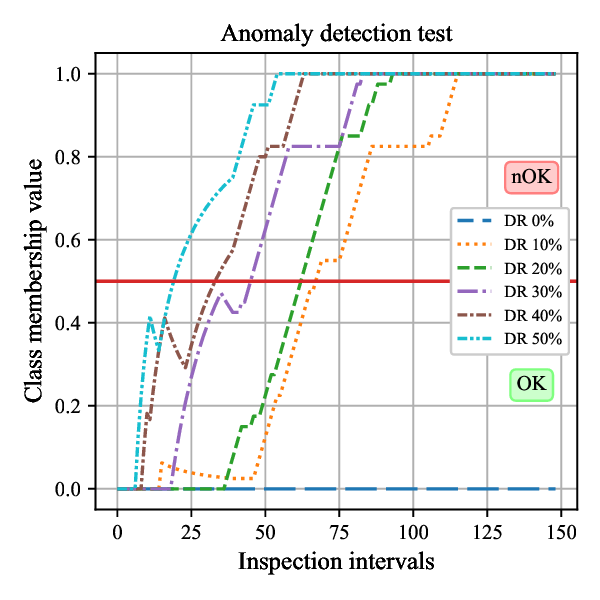}
    \caption{Anomaly identification test on different datastreams}
    \label{fig:anomalyIdent}
\end{figure}

With the intention to implement the proposed algorithm in a live system, an additional anomaly identification test on different datastreams is performed. Therefore an introduction into inspection intervals needs to be done. A predicitve maintenance (PM) framework monitors the health state or condition of a system, represented through parameters and characterstic values, over a given time. Each point of observation in this specific timeline is called an inspection \cite{tinga2020physical}. In the following experiment six different datastreams are tested on the CDF. The reference is represented as a datastream with no degradation at all (DR 0\%), but with existing measurement noise. In addion, five datastreams is applied to the algorithm, each with a different rate of degradation (DR 10\% - DR 50\%). The timeline is set to contain a maximum of $150$ inspections. In the transient states typically an oscillating behavior of the classified class will occure. To counteract this effect, a moving average filter is applied. It contains a sliding window technique (SWT), which is commonly used in PM tasks to capture time dependant information over a certain period of time, in this case $40$ inspections. With the utilization of a filter, a transformation of the discrete classification value into a continuous class membership value is achieved. Therefore a class membership value under $0.5$ is defined as OK with no anomalies and above $0.5$ as not OK (nOK) containing anomalies. The experimental results of the applied datastreams onto the CDF are shown in Fig.~\ref{fig:anomalyIdent}. First, it is noticable that a noise loaded datastream (DR 0\%) of normal operating conditions is correctly classified. Second, every degradation containing datastream is subject to a transmission from state OK to nOK. Third, the higher a degradation rate is the earlier the CDF is able to detect an anomaly. These three aspects indicate that the proposed CDF is detecting anomalies in a robust manner.

\section{Conclusions}
In this paper, a novel change detection framework based on entropy analysis, principle component analysis and fuzzy clustering algorithms is proposed to detect anomalies in time series of pump currents of EDFAs due to degeneration effects. A mentionable characteristic of the proposed method includes integrating multiple sensor data with arbitrary operational conditions in a consolidated framework. The fuzzy clustering based model takes normalized and cleaned sequence data as main inputs and eliminates stochastic non--informant data through entropy analysis. Furthermore, the dimensional reduced data stream is processed by principle component analysis to extract the features with the highest variance ratio and enabling generalization of the given data. Based on experimental measurements of drifting pump currents, the performance of the fuzzy clustering algorithms and the novel change detection framework shows significant generalization under arbitrary operational conditions. This leads to the ability to detect anomalies in the time series with up to 4.9\,\% drift from the typical operating conditions. In addition, the experiments carried out show a robust behavior of the algorithm. In particular, it is possible to robustly detect anomalies in data streams that contain degenerative features as well as noisy data streams that resemble normal operating conditions. Detecting anomalies in time series of data resembles the important initial
\newpage
\noindent step enabling predictive maintenance. Reducing data dimensionality leads to decentralized implementations and a non--traffic--affecting working scenario of such a system. In our future work, the issue of analyzing anomalies in arbitrary operational conditions will be considered to propose a prognostic and diagnostic model for EDFAs.
\vspace*{10pt}

\bibliographystyle{IEEEtran}
\bibliography{IEEEabrv, references.bib}

\end{document}